\documentclass[1p]{elsarticle}
\makeatletter
\def\ps@pprintTitle{%
 \let\@oddhead\@empty
 \let\@evenhead\@empty
 \def\@oddfoot{\centerline{\thepage}}%
 \let\@evenfoot\@oddfoot}
\makeatother
\usepackage{graphicx}
\usepackage{mathptmx}      

\usepackage{amsmath}
\usepackage{enumitem}
\usepackage{subfigure}
\usepackage{multirow}
\usepackage{url}
\usepackage{pifont}
\usepackage{graphicx}
\usepackage{amsmath}
\usepackage{footnote}
\usepackage{algorithm}
\usepackage{algorithmic}
\usepackage{graphicx}
\usepackage{fancyvrb}

\newcommand{\tick}{\ding{52}}

\begin{document}

\title{Inferring Conceptual Relationships when Ranking Patients}


\author[cam]{Nut Limsopatham} 
\ead{nl347@cam.ac.uk}

\author[gla]{Craig Macdonald}
\ead{craig.macdonald@glasgow.ac.uk}

\author[gla]{Iadh Ounis}
\ead{iadh.ounis@glasgow.ac.uk}

\address[cam]{Language Technology Lab, University of Cambridge, United Kingdom}
\address[gla] {School of Computing Science, University of Glasgow, United Kingdom}


%

\begin{abstract}

Searching patients based on the relevance of their medical records is challenging because of the inherent implicit knowledge within the patients' medical records and queries. Such knowledge is known to the medical practitioners but may be hidden from a search system. For example, when searching for the patients with a heart disease, medical practitioners commonly know that patients who are taking the amiodarone medicine are relevant, since this drug is used to combat heart disease. In this article, we argue that leveraging such implicit knowledge improves the retrieval effectiveness, since it provides new evidence to infer the relevance of patients' medical records towards a query. Specifically, built upon existing conceptual representation for both medical records and queries, we proposed a novel expansion of queries that infers additional conceptual relationships from domain-specific resources as well as by extracting informative concepts from the top-ranked patients' medical records. We evaluate the retrieval effectiveness of our proposed approach in the context of the TREC 2011 and 2012 Medical Records track. Our results show the effectiveness of our approach to model the implicit knowledge in patient search, whereby the retrieval performance is significantly improved over both an effective conceptual representation baseline and an existing semantic query expansion baseline. In addition, we provide an analysis of the types of queries that the proposed approach is likely to be effective.

\end{abstract}

\maketitle

\section{Introduction}\label{intro}

\looseness-1 Electronic medical records (EMRs) detail the medical history of patients visiting healthcare providers~\cite{kotsipoulos,tambouris}. Within a patient search system, these EMRs could be exploited to identify cohorts for clinical studies~\cite{limsopatham2013oair,voorhees2011trec,voorhees2012trec}. One of the major challenges of searching in the medical domain is to deal with often complex, inconsistent and ambiguous terminology~\cite{krauthammer2004jbi,limsopatham2011sigir,trieschnigg2010cikm}. For example, it is commonly known by medical practitioners that \emph{`cancer'}, \emph{`carcinoma'}, \emph{`CA'}, and \emph{`malignant tumour'} share a similar meaning. However, such information may be hidden from traditional search systems. To handle such a challenge, prior works resorted to domain-specific resources to improve the representation of medical documents and queries. For instance, by exploiting the knowledge obtained from domain-specific resources, such as MeSH,\footnote{\url{http://www.nlm.nih.gov/mesh/}} using concept-based representation approaches, \emph{`cancer'}, \emph{`carcinoma'}, \emph{`CA'}, and \emph{`malignant tumour'} are represented with the same concept to lessen the mismatch of synonymous terms in medical documents and queries~\cite{trieschnigg2010cikm,hersh1994jamia,srinivasan1996ipma}. Moreover, synonym and hyponym relationships of medical terms obtained from domain-specific resources have been effectively exploited to reformulate queries in order to further improve retrieval performance (e.g.~\cite{aronson1997amia,srinivasan1996ipmb}). Hence, a search system could infer that patients whose medical records stated that they are suffering from \emph{`pulmonary atresia'} are relevant to a query searching for the patients suffering from \emph{`a heart disease'}, since according to MeSH, \emph{`pulmonary atresia'} is a particular form of \emph{`a heart disease'}. 

Nevertheless, the aforementioned approaches do not leverage the inherent implicit knowledge in medical records and queries, which could be exploited to improve retrieval effectiveness. For example, we can infer that a patient is suffering from a particular disease, if the patient is prescribed a medicine used specifically to treat that disease. Statistical query expansion (QE) approaches, such as pseudo-relevance feedback, indirectly deal with the implicit knowledge challenge by using occurrence statistics of terms in the top-ranked documents to improve the representation of the original query~\cite{amati2003thesis}. For example, Limsopatham et al.~\cite{limsopatham2012sigir,limsopatham2015cikm} effectively applied the Divergence from Randomness (DFR) Bo1 QE model~\cite{amati2003thesis} to improve the retrieval performance of a patient search system.
However, these QE approaches might not effectively deal with the implicit knowledge challenge, if inherently related concepts are not observed in the top-ranked records.

\looseness-1 Instead, to leverage the implicit knowledge inherent to the patient search process, in this article, we go beyond classical statistical QE techniques. 
We propose to improve the representation of medical queries by inferring the relationships of medical conditions that are extracted from the queries. In particular, we focus on medical conditions that are related to \emph{four aspects of the medical decision process (namely, symptom, diagnostic test, diagnosis, and treatment)}~\cite{Silfen2006664,limsopatham2013oair,limsopatham2013ecir-b}, as they are important information considered by healthcare practitioners when consulting with a patient.
Next, using this conceptual representation of the medical records and queries, we propose a novel query expansion approach to further improve the query representation by exploiting two types of resources. The first type consists of external resources about the possible relationships between medical conditions (i.e.\ medical concepts). For example, we can infer that a patient suffers from a particular disease, if the patient is prescribed a medicine for that disease. To infer the relationships between medical concepts using external resources, we propose two techniques, which are based on Bayes' theorem and a stochastic analysis, respectively. On the other hand, as a second type of resource, we use the collection of patients' medical records itself. In particular, we extract the most informative concepts from the top-ranked medical records retrieved using the original query. The medical concepts inferred using these two types of resources are used to expand the original queries to improve their representations.
%

We evaluate our proposed approach in the context of the TREC 2011~\cite{voorhees2011trec} and 2012~\cite{voorhees2012trec} Medical Records track. Our results attest the effectiveness of our proposed approach, as it can significantly improve the retrieval performance over an existing query expansion baseline. 

The main contributions of this paper are threefold:
\begin{enumerate}
\item We propose a novel query expansion approach that models the relationships between concepts using the four aspects of medical decision process, by leveraging medical knowledge gained from external resources and the occurrence statistics of concepts in the top-ranked medical records to improve the query representation and to infer relevance.
\item We introduce two techniques for inferring the relationships between medical concepts derived from external resources, based on Bayes' theorem and a stochastic analysis, respectively.
\item We thoroughly evaluate our proposed approach within the standard experimentation paradigm provided by the TREC 2011 and 2012 Medical Records track. 
\end{enumerate}

\looseness-1 The remainder of this paper is organised as follows. In Section~\ref{RelatedWork}, the backgrounds of searching patients based on the relevance of their medical records and related works are discussed. Section~\ref{Approach} discusses our approach for improving query representation that infers relationships between medical concepts by using information from both external resources and the informative concepts extracted from the top-ranked medical records. 
Experimental setup and results are presented in Sections~\ref{c7sSetup} and~\ref{c7sResults}, respectively. Section~\ref{c7sAnalysis} provides a further analysis and discussion of the proposed approach. 
Finally, we conclude the paper in Section~\ref{Conclusions}.

\section{Related Work}\label{RelatedWork}

Electronic medical records (EMRs), which detail the healthcare information of patients, have been developed to improve the quality of healthcare services~\cite{hersh2009bio}. For instance, EMRs could be exploited to identify treatments that have been used effectively to combat a particular disease~\cite{hersh2009bio,hersh2004jama}. However, the characteristics of medical records and queries, such as the complexity of the medical terminology, are different from those of other domains. Hence, effective search approaches for medical records are needed.
In 2011, TREC developed a search task to facilitate the research in this area~\cite{voorhees2011trec}. In particular, the task of the TREC Medical Records track~\cite{voorhees2011trec,voorhees2012trec} aims to rank patients with respect to the relevance of their medical records towards a query. 

Prior work (e.g.~\cite{gurulingappa2011trec,limsopatham2011trec,zhu2012shb}) effectively handled this search task using techniques previously developed for expert search~\cite{balog2008trec},
since the goal of both tasks is to rank people (i.e.\ patients or expert persons) based on the relevance of their associated documents. On one hand, expert search aims to rank experts based on the relevance of the documents they have written or that mention them~\cite{balog2008trec}. On the other hand, patient search uses the estimated relevance of medical records, when ranking their associated patients. Hence, in this work, we also handle patient search using well-establ\-i\-s\-hed approaches previously developed for expert search, 
which use ranked medical records to rank patients (e.g.\ Voting Model~\cite{macdonald2006cikm} and Model~2~\cite{balog2006sigir}). Specifically, the Voting Model ranks patients using a voting process, where the ranking of medical records (denoted $R(Q)$) defines the relevance scores for the patients to be retrieved. Each retrieved medical records in $R(Q)$ is said to vote for the relevance of its associated candidate patient using voting techniques such as, CombMAX, CombSUM, expCombSUM. Indeed, each voting technique firstly ranks medical records based on their relevance towards a query using a traditional weighting model (e.g.\ BM25~\cite{robertson1994sigir}, DFR DPH~\cite{amati2007trec}), and then aggregates the votes from the medical records to their associated patients, to create a ranked list of likely relevant patients for the query~\cite{macdonald2006cikm}. 

One of the important research areas of searching in the medical domain is dealing with
the complexity, ambiguity and inconsistency of the medical terminology~\cite{krauthammer2004jbi,limsopatham2011sigir,trieschnigg2010cikm}. For example, when referring to \emph{`coronary heart disease'}, different medical practitioners may use terms, such as \emph{`coronary artery disease'}, \emph{`arteriosclerotic heart disease'}, \emph{`CHD'}, or \emph{`CAD'}. Previous work has leveraged domain-specific resources to handle such a challenge~\cite{krauthammer2004jbi,limsopatham2011sigir,trieschnigg2010cikm}. For instance, Srinivasan~\cite{srinivasan1996ipma} and Trieschnigg et al.~\cite{trieschnigg2010cikm} represented medical documents and queries using medical concepts obtained from domain-specific resources, such as MeSH, to alleviate the synonymous mismatch of terms in medical documents and queries. Aronson~\cite{aronson1994riao} deployed MetaMap~\cite{aronson2010jamia} -- a medical concept recognition tool based on the UMLS Metathesaurus\footnote{\url{http://www.nlm.nih.gov/research/umls/}} -- to identify all concept, in the medical documents and queries, and to represent them in the form of the UMLS Concept Unique Identifier (CUI). 
However, conceptual representation approaches are effective only when combined with a traditional term-based representation~\cite{trieschnigg2010cikm,srinivasan1996ipma}.
Koopman et al.~\cite{koopman2012adcs} proposed a graph-based approach for weighting patients based on the occurrences of medical concepts in their medical records. They built a graph model based on the co-occurrences of medical concepts within a particular window of concepts, and calculated the PageRank~\cite{brin1998} scores for each medical concepts. Then, during retrieval, they scored patients based on the PageRank scores of the query concepts that matched their medical records. Later, Limsopatham et al.~\cite{limsopatham2013oair} proposed a task-specific representation approach, which represented medical records and queries by using only concepts that are related to four aspects of the medical decision process (namely, symptom, diagnostic test, diagnosis and treatment)~\cite{Silfen2006664}, as they are essential information for health practitioners when dealing with a patient. 
Note that these aspects can be viewed as the criteria aspects that the medical records of relevant patients should cover as suggested in~\cite{koopman2014sigir,limsopatham2014cikm}. 
In this work, we follow~\cite{limsopatham2013oair} when representing medical records and queries. In particular, we deploy MetaMap to extract concepts, in medical records and queries, and represent the identified concepts in the form of the UMLS Concept Unique Identifier (CUI). Importantly, we identify concepts related to the four aforementioned aspects based on their MetaMap's semantic type.\footnote{\url{http://metamap.nlm.nih.gov/SemanticTypeMappings_2011AA.txt}} In Table~\ref{tab:CU}, we list the 16 MetaMap's semantic types that are associated to the medical decision process~\cite{limsopatham2013ecir-b}. Medical concepts with the MetaMap's semantic type defined in the first column are associated to an aspect of the medical decision process, if there is a tick (\tick) in the column of that aspect. For example, concepts having the MetaMap's semantic type \emph{Disease or Syndrome} are associated with the \emph{diagnosis} aspect.
Figure~\ref{fig:c5metamap} shows an example of using the MetaMap tool to extract medical concepts from the query ``Patients with diabetes mellitus who also have thrombocytosis'', where 
we can extract four different CUIs, including `C0011849', `C0011847', `C0241863' and `C0836924', from. Nevertheless, even though this approach could effectively represent medical concepts mentioned in documents and queries, related medical concepts that are not explicitly mentioned may not be inferred and uncovered.

\begin{figure}[tb]
\centering
\begin{Verbatim}[frame=single]
Input: "Patients with diabetes mellitus who also 
	have thrombocytosis"

Phrase: "Patients"

Phrase: "with diabetes mellitus"
Meta Candidates (4):
  1000 C0011849:Diabetes Mellitus [Disease or Syndrome]
          Diabetes
   861 C0011847:Diabetes [Disease or Syndrome]
   789 C0241863:DIABETIC [Finding]
Meta Mapping (1000):
  1000 C0011849:Diabetes Mellitus [Disease or Syndrome]

Phrase: "who also"

Phrase: "have"

Phrase: "thrombocytosis."
Meta Candidates (1):
  1000 C0836924:Thrombocytosis [Disease or Syndrome]
Meta Mapping (1000):
  1000 C0836924:Thrombocytosis [Disease or Syndrome]
\end{Verbatim}
\caption{Medical concepts extracted by the MetaMap tool from the query `Patients with diabetes mellitus who also have thrombocytosis'}\label{fig:c5metamap}
\end{figure}


On the other hand, query expansion (QE) techniques, such as pseudo-relevance feedback, have been used to effectively improve the representation of queries in different search tasks~\cite{aronson1997amia,srinivasan1996ipmb,amati2003thesis}. 
In the form of QE, synonyms and hyponyms of concepts in the medical documents and queries have been used to improve the representation of medical queries
~\cite{aronson1997amia,martinez2014,srinivasan1996ipmb,zuccon2012}. For example, Aronson and Rindflesch~\cite{aronson1997amia}, and Srinivasan~\cite{srinivasan1996ipmb} effectively expanded concepts in a query with their synonyms and hyponyms obtained from domain-specific resources, such as MeSH and UMLS Metathesaurus. In the context of patient search, King et al.~\cite{king2011trec} effectively deployed the UMLS Metathesaurus and their collection of medical reference encyclopedias in their query expansion technique. In particular, they simply added the derived medical concepts to the query and set the weights equally for all the medical concepts. Zuccon et al.~\cite{zuccon2012} investigated into exploiting hierarchical structure of SNOMED-CT ontology\footnote{\url{https://www.nlm.nih.gov/snomed/}} to extract expansion concepts and several techniques (e.g. tf-idf) for calculating weights of the expansion concepts. However, they found a limited performance improvement. Recently, Martinez et al.~\cite{martinez2014} achieved significant improvement when applied PageRank random walks on the UMLS Metathesaurus to extract expansion concepts and their weight. However, a limitation is that some types of relationships between medical concepts (e.g.\ the relationships between diagnosis and treatment) may not be directly presented in the UMLS Metathesaurus.
Meanwhile, pseudo-relevance feedback is to expand a query with \emph{a set of informative terms} obtained from the top-ranked documents. In the context of searching the medical domain, Srinivasan~\cite{srinivasan1996ipmb} and Limsopatham et al.~\cite{limsopatham2012sigir} reported that pseudo-relevance feedback could effectively improve retrieval performance. In particular, pseudo-relevance feedback can indirectly deal with the implicit knowledge, since it may expand the query with semantically-related concepts. For example, for a query searching for \emph{``patients with heart disease"}, the expanded terms might be \emph{`amiodarone'} and \emph{`angina'}, which are a treatment and a symptom associated to \emph{`heart disease'}. Importantly, in medical records such relationships between concepts related to the aspects of the medical decision process are strongly established. On the other hand, from a medical record of a patient taking \emph{`amiodarone'} medicine (treatment), healthcare practitioners can infer that the patient is suffering from \emph{`heart disease'} (diagnosis), since according to the indication \emph{`amiodarone'} is used to combat \emph{`heart disease'}. However, pseudo-relevance feedback may not always be able to leverage such implicit knowledge, if the associated concepts do not appear in the top-ranked medical records. Hence, in this work, we propose to also directly infer these relationships to improve the query representation. Specifically, we propose a new QE approach that uses conceptual relationships extracted from both external resources and the top-ranked medical records to improve retrieval effectiveness.

\begin{table*}[tb]
  \caption{\looseness-1  List of 16 of the MetaMap's 133 semantic types that we consider for our proposed approach, based on the four aspects of the medical decision process.}
  \label{tab:CU}
  \centering 
\resizebox{1.0\textwidth}{!}{
\begin{tabular}{|l|c|c|c|c|}
\hline
\multirow{2}{*}{MetaMap's Semantic Type} & \multicolumn{4}{c|}{Aspects of the Medical Decision Process}\\ \cline{2-5}
& Symptom & Diagnostic test & Diagnosis & Treatment \\
\hline
Body Location or Region & \tick & \tick & \tick & \tick \\
Body Part, Organ, or Organ Component  & \tick & \tick & \tick & \tick \\
Clinical Drug & -- & -- & -- & \tick \\
Diagnostic Procedure& -- & \tick & -- & --\\
Disease or Syndrome&  -- & -- & \tick & --\\
Finding & \tick & -- & -- & --\\ 
Health Care Activity&  -- & \tick & -- & \tick \\
Injury or Poisoning& \tick & -- & -- & --\\
Intellectual Product& -- & \tick & -- & \tick \\
Medical Device & -- & \tick & -- & \tick \\
Mental or Behavioral Dysfunction & \tick & -- & \tick & --\\
Neoplastic Process& \tick & \tick & \tick & \tick \\
Pathologic Function& \tick & -- & -- & --\\
Pharmacologic Substance & -- & -- & -- & \tick \\
Sign or Symptom& \tick & -- & -- & --\\
Therapeutic or Preventive Procedure & -- & -- & -- & \tick \\
\hline
\end{tabular}
}
\end{table*}
%
%

\section{An Approach for Inferring Conceptual Relationships when Ranking Patients}\label{Approach}
In this section, we introduce our proposed approach for inferring the relationships between medical concepts that leverage both external resources and the local statistics of the top-ranked medical records. In particular, Section~\ref{c7sRules} discusses how we derive possible relationships between medical concepts from external resources and represent them as association rules. In Section~\ref{All} and~\ref{c7sRandomWalk}, we propose two techniques based on Bayes' theorem and a stochastic analysis, respectively, to infer relationships between medical concepts from the extracted association rules. Section~\ref{c7sModel} discusses our approach to improve the representation of the query by using both medical concepts derived from the external resources using either Bayes' theorem or a stochastic analysis, and informative concepts extracted from the top-ranked medical records.

\subsection{Extracting and Representing Medical Concept Relationships from External Resources as Association Rules}\label{c7sRules}

Driven by the aforementioned medical decision process~\cite{limsopatham2013ecir-b}, we extract directed association rules representing the relationships between concepts from two different types of medical resources\footnote{We provide the list of all of the used resources in Section~\ref{c7sSetup}.}, which are ontology-based and free-text-based resources, respectively. We use different strategies for extracting conceptual relationships from each type of resource. For the ontology-based resources (e.g.\ MedDRA\footnote{\url{http://www.meddramsso.com}} and DOID\footnote{\url{http://purl.bioontology.org/ontology/DOID}}), we use the semantic relationships of concepts within each ontology to represent the relationships between concepts. For instance, `Coronary heart disease' is a particular type of `heart disease' (i.e.\ \emph{`Coronary heart disease'$\rightarrow$`heart disease'} -- A patient suffering from `Coronary heart disease' can be considered as having `heart disease' condition). 

\begin{figure}[tb]
  \centering
\includegraphics[angle=0,width=100mm]{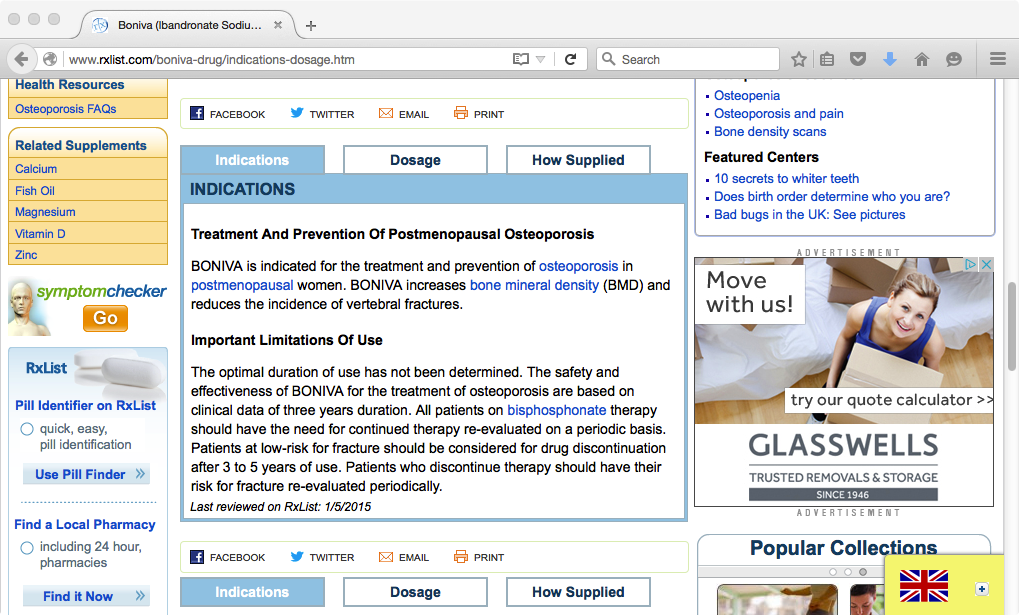}
\caption{An example of the indications of `Boniva' from \protect\url{http://www.rxlist.com} website.}\label{fig:rxlist}
\end{figure}

For the free-text-based resources (e.g.\ \url{http://www.rxlist.com}), we use Meta\-Map to identify concepts related to the four aforementioned medical aspects from the free-text documents (we use the task-specific representation approach~\cite{limsopatham2013ecir-b} described in Section~\ref{RelatedWork} to identify medical concepts from sentences in the free-text documents), and then assume the existence of relationships between the identified concepts.
For example, from a drug indication in the \url{http://www.rxlist.com} website, which states that \emph{``Boniva (ibandronate sodium) is indicated for the treatment and prevention of osteoporosis in postmenopausal women''}, as shown in Figure~\ref{fig:rxlist},
MetaMap can identify the concepts \emph{`Boniva'} (treatment) and \emph{`osteoporosis'} (diagnosis). Assuming the relationships between medical concepts found in the drug description, we surmise that there is an association between the two concepts (i.e.\ \emph{`Boniva'$\rightarrow$`osteoporosis'} and \emph{`osteoporosis'$\rightarrow$`Boniva'}). Next, the extracted association rules are stored in a database. For instance, as shown in Table~\ref{tab:reasoning}, the rules associated to the concept \emph{`osteoporosis'} are \emph{`Dowager's hump'$\rightarrow$`os\-teoporosis'}, \emph{`DEXA'$\rightarrow$`os\-teoporosis'}, \emph{`Prolia'}\emph{$\rightarrow$`os\-teo\-porosis'}, and \emph{`Boniva'$\rightarrow$`os\-teo\-poro\-sis'}.
These association rules provide new evidence to infer the possible relevance of medical records, which are then aggregated to estimate the relevance of their associated patients. For instance, we can infer that patients taking the \emph{`Boniva'} medicine suffer from \emph{`osteoporosis'}, since \emph{`Boniva'} is a treatment for \emph{`osteoporosis'}.

\begin{table}[tb]
\caption{Examples of medical concepts associated to `osteoporosis' defined in our associated rules database.}\label{tab:reasoning}
  \centering
\begin{tabular}{|l|l|}
    \hline
    Medical Concept & Related Concept\\ \hline
    Dowager's hump & Osteoporosis\\ 
    DEXA & Osteoporosis\\ 
    Prolia & Osteoporosis\\ 
    Boniva & Osteoporosis\\ \hline
  \end{tabular}   
\end{table}


\subsection{A Bayesian-based Technique for Inferring Conceptual Relationships}\label{All}
\looseness-1 We argue that the relationships between concepts that are related to the medical decision process could be leveraged to deal with the complexity of medical terminology.
For example, if we have evidence that a patient is taking the \emph{`olmesartan'} medicine (treatment), we can infer that the patient suffers from \emph{`hypertension'} (diagnosis), since \emph{`olmesartan'} is a treatment for \emph{`hypertension'}. Therefore, using external domain-specific resources, we propose to 
reformulate the queries by using the association rules extracted from the medical resources (described in Section~\ref{c7sRules}). 

Specifically, we first retrieve a set of candidate concept expansions (denoted $inferred(q)$) corresponding to the query concepts from the extracted association rules. Then, to prevent excessively general candidate concepts being added to the query, we estimate the association of a query concept and each candidate concept expansion using a Bayesian probabilistic score computed based on the occurrences of concepts in the association rules (both derived from ontologies and free-text resources). The higher the probability, the stronger the relationship between the two concepts. Indeed, the Bayesian probabilistic score of the association between query concept $t$ and its corresponding concept $t'$ is estimated as follows:
\begin{align}\label{eq:conditional}
w_a(t,t') = p(t'|t) = \frac{p(t' \cap t)}{p(t)}
\end{align}
where $p(t' \cap t)$ is the maximum likelihood that the concept $t'$ co-occurs with the query concept $t$ within all the extracted association rules, and $p(t)$ is the maximum likelihood that the concept $t$ is contained in an association rule. 
This maximum likelihood is calculated based on the frequency of the occurrences of each concept in the whole association rules database.

\begin{figure}[tb]
  \centering
\includegraphics[angle=0,width=85mm]{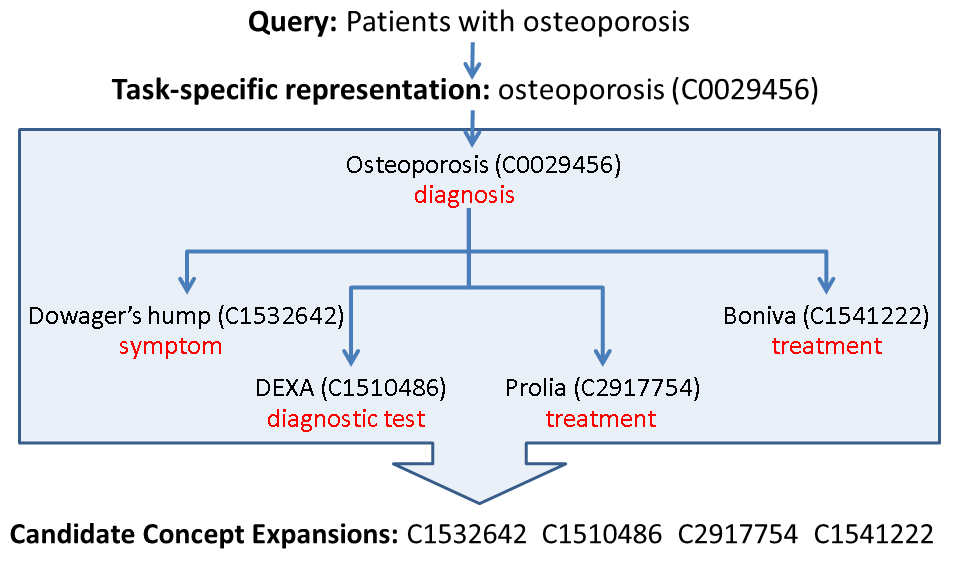}
\caption{An example of identifying candidate con\-cept expansions using our technique on query \emph{``patients with osteoporosis"}.}\label{fig:reasoning}
\end{figure}

Figure~\ref{fig:reasoning} shows an example of how our technique identifies candidate concept expansions for the query \emph{``patients with osteoporosis"}. In particular, we first obtain the concept \emph{`osteoporosis'} from the query using the task-specific representation approach. Then, the concept \emph{`osteoporosis'} is used to retrieve related concepts from the database of association rules previously extracted from the medical resources (see Section~\ref{c7sRules}). As shown in Figure~\ref{fig:reasoning}, the retrieved candidate concept expansions include \emph{`Dowager's hump'}, \emph{`DE\-XA'}, \emph{`Prolia'} and \emph{`Boniva'}, which are the symptom, diagnostic test, and treatments associated to the original query concept (i.e.\ \emph{`osteoporosis'}). 
However, as previously discussed in Section~\ref{RelatedWork}, QE is effective when a number of highly informative terms that are associated to the query are added to the query~\cite{amati2003thesis}, we select a top-k candidate concept expansions, which are ranked based on the score computed using Equation~(\ref{eq:conditional}), to expand the original query. 

\subsection{A Stochastic Analysis for Inferring Conceptual Relationships}\label{c7sRandomWalk}

Next, we discuss our stochastic analysis technique for inferring relationships between medical concepts in the association rule database built in Section~\ref{c7sRules}. Indeed, we propose to measure the strength of the relationships between the medical concepts in the association rule database using a stochastic analysis of some random walk behaviour through the association rules.

\begin{figure*}[tb]
\centering
        \subfigure[A collection \emph{D} of documents \emph{1, 2, 3, 4, 5}]{
            \label{fig:c7:sub-a}
            \includegraphics[angle=0,width=70mm,scale=1]{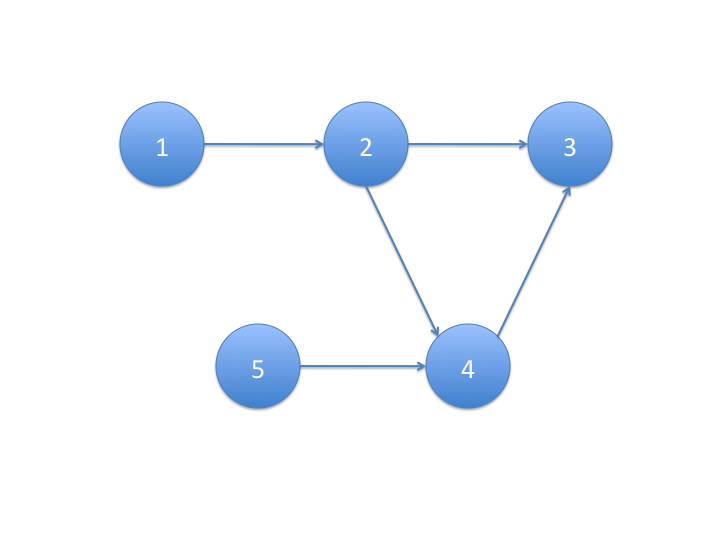}
        }
        \subfigure[A bipartite graph \emph{G}]{
           \label{fig:c7:sub-b}
           \includegraphics[angle=0,width=70mm,scale=1]{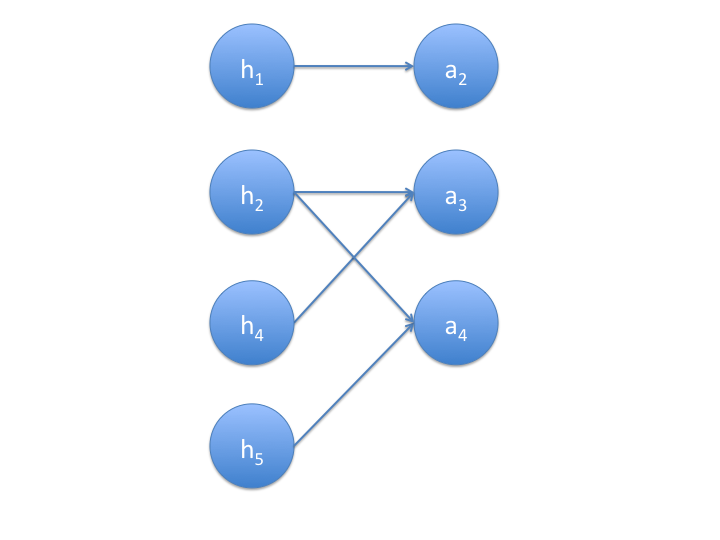}
        }
    \caption{%
        Transforming (a) the collection \emph{D} into (b) a bipartite graph \emph{G}.
     }%
   \label{fig:example}
\end{figure*}

A stochastic analysis of random-walk behaviour has been studied before in IR (e.g.~\cite{brin1998,lempel2001tois,blanco2012ir}). For example, \cite{brin1998}'s PageRank used a stochastic analysis of random walks on the hyperlink of the entire Web to determine the importance of each web page. \cite{lempel2001tois} introduced a stochastic approach for link-structure analysis (SALSA), which performed random walks on a sub-graph derived from the link-structure of an initial set of retrieved results. 
In particular, consider a collection \emph{D} of documents (e.g.\ \emph{1, 2, 3, 4 and 5}) and directed relationships between the documents (see Figure~\ref{fig:c7:sub-a}). For example, there is a relationship from document $1$ to document $2$. The approach first converts the collection \emph{D} to a bipartite graph \emph{G}, of which the two parts are considered \emph{hubs} and \emph{authorities}. Hubs are documents (i.e.\ nodes) that link to other documents, while authorities are documents that are linked to by other documents. The SALSA algorithm performs random walks to uncover the relationships between these documents. 

Different from existing work, we propose to apply the SALSA algorithm to uncover medical concepts that are related to a given query and identify the strength of their relationships, using the link structure within our association rule database. Later, we discuss in Section~\ref{c7sModel} how these medical concepts are used to enrich the query.

Specifically, we adapt the SALSA algorithm to perform the analysis of the links between the concepts in the association rules. In this work, the nodes (e.g.\ \emph{1, 2, 3, 4 and 5} in Figure~\ref{fig:c7:sub-a}) are the medical concepts in the association rule database, while an edge is a directed relationship between two concepts (i.e.\ nodes). Consider a bipartite graph \emph{G} where the two parts are hubs and authorities, a directed edge between hub $h$ and authority $a$ is indicated by a conceptual relationship between them (e.g.\ the edge between concept \emph{1} and concept \emph{2} -- $h_1 \rightarrow a_2$ in Figure~\ref{fig:c7:sub-b}). A good hub points to many authorities, and a good authority is pointed by many hubs. 

The authoritative medical concepts related to the query $q$ should be linked to by many medical concepts in the sub-graph induced by a set of medical concepts $C$ in the query $q$. A random walk on the sub-graph induced by $C$ will visit those authoritative concepts with high probability. Hence, our adaptation of the SALSA algorithm~\cite{lempel2001tois} identifies medical concepts that are related to those medical concepts occurring in the query $q$, by using the iterative algorithm presented in Algorithm~\ref{al:c7:salsa}.

The medical concepts with high authority scores are the concepts highly related to the query $q$, as they can be inferred by many concepts related to the query. 
To incorporate these medical concepts while ranking patients based on the relevance of their medical records, we select top-k candidate concept expansions based on their authoritative scores computed using Algorithm~\ref{al:c7:salsa}.

\begin{algorithm}[tb]
\begin{algorithmic}[1]
\STATE Initialise $a(s) \leftarrow 1$, $h(s) \leftarrow 1$ for all concepts $s \in C \cup C_a$, where $C_a$ is a set of medical concepts in the association rule database that link to or are linked to by a concept in $C$.
\STATE Repeat the following steps until convergence:
\STATE ~~~~~~~~Update the authority score of each concept $s$:
\STATE ~~~~~~~~~~~~~~~~$a(s) \leftarrow \sum_{\{ x | x \textrm{ points to } s\}} h(x)$
\STATE ~~~~~~~~Update the hub score of each concept $s$:
\STATE ~~~~~~~~~~~~~~~~$h(s) \leftarrow \sum_{\{ x | s \textrm{ points to } x\}} a(x)$
\STATE ~~~~~~~~Normalise the authority scores and the hub scores.
\end{algorithmic}
\caption{The Adapted SALSA Algorithm}
\label{al:c7:salsa}
\end{algorithm}

\subsection{Modelling Conceptual Relationships when Ranking Medical Records}\label{c7sModel}
We leverage both the local statistics from the top-ranked medical records and the association rules extracted from external resources to improve the representation of a given query. 
In particular, we apply a pseudo-relevance feedback technique to identify informative concepts from the top-ranked medical records retrieved using the original query. 
Indeed, when applying a pseudo-relevance feedback technique, concepts occurring in the top-ranked medical records are firstly weighted and ranked using a term weighting model. Then, the top-ranked concepts (i.e.\ the most informative concepts) are used to expand the original query. 
For example, for the query \emph{``patients with vascular disease"}, which is represented as \emph{``C0042373"} (\emph{`vascular disease'}), a pseudo-relevance feedback QE technique could identify related concepts, such as \emph{`C0190932'} (\emph{`femoral-popliteal artery bypass graft'}) and \emph{`C0014098'} (\emph{`endarterectomy'}), which correspond to treatments for that disease. However, the related diagnostic procedures and symptoms, such as \emph{`C0202896'} (\emph{`carotid angiogram'}), might not be added to the query, if they do not appear in the top-ranked medical records. 
Hence, on other hand, we also expand the query with medical concepts inferred using the association rules. We identify candidate expansion concepts and their weights using either the Bayesian-based technique or the stochastic technique introduced in Section~\ref{All} or~\ref{c7sRandomWalk}, respectively.

In particular, our approach estimates the relevance score of a medical record $d$ towards the query $q$ as follows:
\begin{align}\label{eq:combining}
score(d,q)& = \sum_{t'' \in q_e} qtw(t'') \cdot score(d,t'') \\ \nonumber
&+ \lambda_r \cdot \sum_{t \in q}\sum_{t' \in inferred(q)} w_a(t,t') \cdot score(d,t')
\end{align}
where $t''$ and $qtw(t'')$ are a medical concept and its term weight in the expanded query $q_e$, which is reformulated using the occurrence statistics of medical concepts in the top-ranked medical records using any pseudo-relevance feedback model (e.g.\ Bo1 from the Divergence from Randomness (DFR) framework). $score()$ can be calculated using any term weighting model such as BM25,
$inferred(q)$ returns a set of expanded concepts, which are related to the concepts in the original query $q$. We use either of the techniques previously proposed in Section~\ref{All} or~\ref{c7sRandomWalk} to extract the expanded concepts (i.e.\ $t' \in inferred(q)$) and their weights (i.e.\ $w_a()$) from the association rule database. In particular, the Bayesian-based technique, and the stochastic analysis technique uses Equation~(\ref{eq:conditional}) and the authority score ($a()$) computed using Algorithm~\ref{al:c7:salsa}, respectively, to calculate $w_a()$. $\lambda_r$ is a parameter to weight the importance of the relevance scores computed for the medical concepts derived using either of the two proposed techniques.

To learn the $\lambda_r$ value for each query, we use the Gradient Boosted Regression Trees (GBRT) regression technique~\cite{tyree2011www} (as implemented in the jforest package~\cite{ganjisaffar2011sigir}\footnote{\url{http://code.google.com/p/jforests}}), since it has shown to be effective for regression tasks (e.g.~\cite{limsopatham2013sigir}). To learn an value of parameter $\lambda_r$ for an unseen query, we use 12 features, which measure the predicted difficulty of the query. An effective feature should indicate the level of emphasis on the relevance scores computed for the concepts inferred using the association rules. Our intuition for using these features is that if the query is difficult, then it might be beneficial to take into account the relevance estimated from the inferred concepts. In particular, if using only the information from the original query is difficult for a search system, medical concepts inferred using the extracted association rules might bring novel evidence that can improve the representation of the query and hence enhance the retrieval performance. 
Table~\ref{tab:features-c7} lists the 12 query performance predictors computed on the original query, which are well-known for measuring the difficulty of a query. 
Specifically, the first set of features (Features 1-4), including the clarity score~\cite{cronen-townsend2002sigir}, SCQ~\cite{zhao2008ecir}, MaxSCQ~\cite{zhao2008ecir} and NSCQ~\cite{zhao2008ecir}, consider the ambiguity of a query by measuring the coherence of the language used in each medical record. 
The more similar the query model is to the collection model, the better the retrieval performance would be expected. The next set of features (Features 5-8) measure the specificity of each query. Indeed, queries with explicit intents could result in a better performance than queries with general concepts. The features include Average Inverse Collection Term Frequency (AvICTF)~\cite{carmel2010sl}, Average Inverse Document Frequency (AvIDF)~\cite{carmel2010sl}, EnIDF~\cite{carmel2010sl}, and the query scope ($\omega$)~\cite{he2006is}. Features 9-10 measure the distribution of informativeness among the query concepts (i.e.\ $\gamma_1$ and $\gamma_2$~\cite{he2006is}), as a query with informative concepts could attain an effective retrieval performance. Next, Feature 11, the Average of the Pointwise Mutual Information over all query term pairs (AvPMI)~\cite{carmel2010sl}, focuses on the relationship between query concepts. The more co-occurrences among the query concepts, the better the chance that the relevant documents are being retrieved.
Finally, Feature 12 is the number of non-stopword query concepts.

\begin{table}[tb]
\caption{List of the features used for learning the emphasise on medical concepts derived from association rules.}
\label{tab:features-c7}   
\centering
\begin{tabular}{|r|l|}
    \hline
    ID & Feature\\ \hline
    1&Clarity Score~\cite{cronen-townsend2002sigir} \\ \hline
    2&SCQ~\cite{zhao2008ecir} \\ \hline
    3&MAXCQ~\cite{zhao2008ecir} \\ \hline
    4&NSCQ~\cite{zhao2008ecir} \\ \hline
    5&AvICTF~\cite{carmel2010sl} \\ \hline
    6&AvIDF~\cite{carmel2010sl} \\ \hline
    7&EnIDF~\cite{carmel2010sl} \\ \hline
    8&Query Scope ($\omega$)~\cite{he2006is} \\ \hline
    9&$\gamma_1$~\cite{he2006is} \\ \hline
    10&$\gamma_2$~\cite{he2006is} \\ \hline
    11&AvPMI~\cite{carmel2010sl} \\ \hline  
    12&Query length~\cite{he2006is}\\ \hline
  \end{tabular}   
\end{table}

\section{Experimental Setup}\label{c7sSetup}

\subsection{Corpus/Queries/Measures}
We evaluate our proposed approach in the context of the TREC 2011~\cite{voorhees2011trec} and 2012~\cite{voorhees2012trec} Medical Records track.
In this track, the task is to identify relevant patient \emph{visits} for a given query topic. Each visit contains all of the medical records associated with a patient's visit to a hospital. A visit is used to represent a \emph{patient} as a unit of retrieval, due to privacy concerns~\cite{voorhees2011trec}. The medical records collection consists of approximately 102k medical records, which can be mapped to 17,265 patient visits, from the University of Pittsburgh NLP Repository. We evaluate our proposed approach using the 34 and 47 topics from the TREC 2011 and 2012 Medical Records track, respectively.

We evaluate the retrieval effectiveness of our proposed approach, in terms of the TREC official measures, which are bpref measure~\cite{buckley2004sigir} for TREC 2011, and infNDCG measure~\cite{yilmaz2008sigir} for TREC 2012. In particular, bpref is the official measure of TREC 2011, since the absolute number of judged patient visits per topic is relatively small~\cite{voorhees2011trec}. bpref is designed for evaluating environments with incomplete relevance data and penalises a system which ranks a judged non-relevant document above a judged relevant document~\cite{buckley2004sigir}. On the other hand, infNDCG is the primary measure of the TREC 2012 Medical Records track~\cite{voorhees2012trec}, since the gold standard judgements are incomplete; hence a sampling approach is deployed to infer the NDCG performance. In addition, we measure statistically significant differences between the retrieval performance achieved by our approach and the baselines using the paired t-test at $p < 0.05$.

\subsection{Medical Record \& Patient Ranking}
\looseness-1 We index the medical records using the Terrier retrieval platform~\cite{ounis06terrier-osir}. The BM25 weighting model~\cite{robertson1994sigir} is used to rank medical records (e.g.\ $score_t()$ in Equation~(\ref{eq:combining})).
Then, to rank the patients, as explained in Section~\ref{RelatedWork}, we deploy the expCombSUM 
voting technique~\cite{macdonald2006cikm}, which gives more importance to the highly relevant medical records while voting for the relevance of the patients. 
In particular, for a given ranking of medical records ($R(Q)$) with respect to query $Q$, each medical record is said to vote for the relevance of its associated patient. The number of medical records in $R(Q)$ to vote for the relevance of the patients is limited to 5,000, as suggested in~\cite{limsopatham2011trec}.


\subsection{Association Rules}

To create the association rule database (discussed in Section~\ref{c7sRules}), we use the external resources listed in Table~\ref{tab:resources}, which are representatives of both ontology-based and free-text-based medical resources. Table~\ref{tab:resources:stat} shows the number of association rules between concepts extracted from the 7 domain-specific resources. The types of relationships that are extracted from each resource are described in Column 2 of Table~\ref{tab:resources}. Note that there are some association rules that overlap between resources. In total, there are 11,373,014 extracted association rules in our database. Meanwhile, Table~\ref{tab:rule:count} shows the number of the extracted association rules with respect to the four aspects of the medical decision process. For example, the rules with the type `diagnosis $\rightarrow$ symptom' are the association rules that can be used to infer the related symptoms that the medical records of the relevant patients are likely to contain if the query contains a particular diagnosis. Note that there are some association rules that are duplicated among different types of rules, since using the technique described in~\cite{limsopatham2013ecir-b}, some medical concepts can be categorised into several medical aspects. For example, medical concepts related to parts of body are considered to be associated with symptom, diagnostic test, diagnosis and treatment (see Table~\ref{tab:CU}). We use these extracted association rules to extract concept expansions by using the Bayesian-based and stochastic techniques previously introduced in Sections~\ref{All} and~\ref{c7sRandomWalk}, respectively. In addition, we follow~\cite{amati2003thesis} and use only \emph{top 10} candidate concept expansions ($k=10$) extracted using each technique.

\begin{table*}[tb]
\caption{List of resources used for extracting the conceptual relationships related to the four aspects of the medical decision process.}\label{tab:resources}
  \centering
\resizebox{1.0\textwidth}{!}{
  \begin{tabular}{|l|p{8.5cm}|}
    \hline
    Resource & Description of the Extracted Association Rules\\ 
    \hline
    DOID hierarchy & Hierarchical relationships between concepts within the same aspects e.g.\ a general disease and a specific type of the general disease\\ 
    MeSH & Hierarchical relationships between concepts within the same aspects e.g.\ a general disease and a specific type of the general disease\\ 
    MedDRA & Hierarchical relationships between concepts within the same aspects e.g.\ a general disease and a specific type of the general disease\\
    DOID & Relationships between concepts across the aspects e.g.\ a disease and its symptoms\\
    \url{http://www.rxlist.com} & Relationships between concepts across the aspects e.g.\ a medicine and the diseases that it can remedy\\
    \url{http://www.webmd.com} & Relationships between concepts across the aspects e.g.\ a diagnostic test and the diseases that it can diagnosed\\
    UMLS & Both hierarchical and across-aspects relationships e.g.\ a general disease and a specific type of the general disease\\
    \hline
  \end{tabular}
}
\end{table*}

\begin{table}[tb]
\caption{Number of association rules extracted from each domain-specific resource.}\label{tab:resources:stat}
  \centering
  \begin{tabular}{|l|l|r|}
    \hline
    Resources & Association Types & \# of Rules\\ 
    \hline
    DOID hierarchy & Specific-general& 2,046\\
    MeSH & Specific-general& 53,915\\
    MedDRA & Specific-general & 86,109\\
    DOID & Across aspects& 9,664\\
    \url{http://www.rxlist.com} & Across aspects & 5,433\\
    \url{http://www.webmd.com} & Across aspects& 3,694\\
    UMLS & Specific-general \& Across aspects & 11,212,153\\
    \hline
  \end{tabular}
\end{table}

\begin{table}[tb]
\caption{The types of the extracted association rules in terms of the four medical aspects.}\label{tab:rule:count}
  \centering
\begin{tabular}{|l|l|}
    \hline
Rule type & \# of rules \\ \hline
diagnosis $\rightarrow$ diagnosis&2,900,788\\
diagnosis $\rightarrow$ diagnostic test&1,907,492\\
diagnosis $\rightarrow$ symptom&2,306,712\\
diagnosis $\rightarrow$ treatment&2,018,061\\
diagnostic test $\rightarrow$ diagnosis&1,905,521\\
diagnostic test $\rightarrow$ diagnostic test&2,726,341\\
diagnostic test $\rightarrow$ symptom&2,023,443\\
diagnostic test $\rightarrow$ treatment&2,695,757\\
symptom $\rightarrow$ diagnosis&2,303,486\\
symptom $\rightarrow$ diagnostic test&2,020,294\\
symptom $\rightarrow$ symptom&4,205,071\\
symptom $\rightarrow$ treatment&2,096,350\\
treatment $\rightarrow$ diagnosis&2,015,362\\
treatment $\rightarrow$ diagnostic test&2,698,453\\
treatment $\rightarrow$ symptom&2,097,231\\
treatment $\rightarrow$ treatment&6,597,028\\
\hline
  \end{tabular}   
\end{table}

For the local-statistic query expansion technique used in Equation~(\ref{eq:combining}) of Section~\ref{c7sModel}, we deploy a parameter-free Bose-Einstein statistics-based (Bo1) model from the DFR framework to extract informative concepts from the top-ranked medical records as it has shown to be effective in previous works~\cite{limsopatham2011trec,limsopatham2013ecir-b}. DFR Bo1 measures the informativeness of a concept based on the divergence of the concept's distribution in the pseudo-relevant feedback documents (i.e.\ top-ranked medical records) from its distribution in the entire collection. The more the divergence is measured, the more likely that the concept is informative and related to the query. Specifically, Bo1 from the Divergence from Randomness framework calculates the weight $w(t)$ of a concept $t$ in the pseudo-relevant documents, as follows~\cite{amati2003thesis}:
\begin{equation}
w(t) = tf_x \cdot log_2 \frac {1+P_n}{P_n} + log_2 (1+P_n)
\end{equation}
where $tf_x$ is the frequency of concept $t$ in the set of pseudo-relevant documents $x$. $P_n = \frac{TF}{N}$, where $TF$ is the term frequency of $t$ in all documents in the collection, and $N$ is the number of documents in the collection. In this work, we follow~\cite{amati2003thesis} and set the number of pseudo-relevant documents and expanded concepts to 3 and 10, respectively.

Once we identify the expanded concepts, a parameter-free function is used to calculate the query term weight $qtw(t)$ of each concept $t$ in the expanded query, as follows~\cite{amati2003thesis}:
\begin{align}
qtw(t) &= \frac{qtf}{qtf_{max}} + \frac{w(t)}{\lim_{TF \rightarrow tfw} w(t)} \\ \nonumber
&= TF_{max} \log_2 \frac{1+P_{n,max}}{P_{n,max}} + \log_2 (1+P_{n,max})
\end{align}\label{eq:c2sBo12}
where $qtf$ is the frequency of a concept $t$ in the query, and $qtf_{max}$ is the maximum of $qtf$. $\lim_{TF \rightarrow tfw} w(t)$ is the upper bound of $w(t)$, $TF_{max}$ is the frequency $TF$ of the concept with the maximum $w(t)$ in the pseudo-relevant documents. $P_{n,max} = \frac{TF_{max}}{N}$. Note that if an original query concept $t$ does not appear in the most informative concept list extracted from the pseudo-relevant documents, its query term weight $qtw(t)$ remains equal to the original one.

As both the Bayesian-based and the stochastic analysis-based techniques require the $\lambda_r$ parameter in Equation~(\ref{eq:combining}) to be properly set, we apply two different training regimes. 
Firstly, we apply a five-fold cross-validation (5-fold) within each set of queries, since the target measure of TREC 2011 and 2012 queries are different. Secondly, we use the TREC 2012 queries as training data, when testing with the TREC 2011 queries, and vice versa. We refer to this setting as \emph{cross-collection validation} (x-collection). 
We optimise the TREC primary measure (i.e.\ bpref and infNDCG measures for TREC 2011 and 2012, respectively) when learning the $\lambda_r$ parameter.

\subsection{Baselines}
We compare the retrieval effectiveness of the proposed approach to infer relationships between medical concepts with the following baselines:
\begin{itemize}
\item The task-specific representation baseline~\cite{limsopatham2013ecir-b} (see Section~\ref{RelatedWork}), which does not apply any QE technique.
\item The Bo1 baseline: applying the Bo1 query expansion technique with the task-specific representation approach.
\item The Bo1 \& Semantic QE baseline: using both the Bo1 query expansion technique and a semantic query expansion technique that adds all of the related medical concepts into the queries (as used in~\cite{king2011trec}). Note that this baseline uses the same set of external resources as the proposed approach.
\end{itemize}

\section{Experimental Results}\label{c7sResults}

\begin{table}[tb]
\caption{Retrieval performance of our Bayesian-based approache for inferring relationships between medical concepts in comparison with the three baselines on TREC 2011 and 2012 Medical Records track's queries. Statistical significance (paired t-test) at $p<0.05$ over the task-specific representation and the Bo1 \& Semantic QE baselines are denoted $^t$ and $^{s}$, respectively.}\label{tab:c7Bayesian}
  \centering
\resizebox{1.0\textwidth}{!}{
  \begin{tabular}{|l|c|c|}
    \hline
     Approach & bpref (TREC 2011) & infNDCG (TREC 2012) \\ 
    \hline
     Task-specific representation & 0.5243 & 0.4880\\
     + Bo1 & 0.5403 & 0.5129 \\
     + Bo1 \& Semantic QE & 0.3013 & 0.2228\\
     + Our Bayesian-based approach (5-fold) & {\bf0.5474}$^{s}$ & {\bf0.5133}$^{t,s}$ \\
     + Our Bayesian-based approach (x-collection) & 0.5464$^{s}$ & 0.5125$^{t,s}$\\
\hline
     + Our Bayesian-based approach (oracle) & 0.5646$^{t,s}$ & 0.5393$^{t,s}$ \\
    \hline
  \end{tabular}
}
\end{table}

In Sections~\ref{c7sExperimentBayesian} and~\ref{c7sExperimentRandomWalk}, we evaluate the proposed approach when using Bayes' theorem and a stochastic analysis to derive associated medical concepts $t''$ and their weights $qtw(t'')$ in Equation~(\ref{eq:combining}), respectively. In particular, we refer the two variants as the Bayesian-based approach and the stochastic approach.

\subsection{Experiment with our Bayesian-based Approach for Inferring Conceptual Relationships}\label{c7sExperimentBayesian}

We first evaluate the effectiveness of the Bayesian-based technique for inferring the relationships between medical concepts. 
To have a fair train/test setting, we use the \emph{cross-collection validation} and \emph{5-fold cross validation} as described in Section~\ref{c7sSetup} when setting $\lambda_r$ within the Bayesian-based approach. Furthermore, to see how the parameter setting impacts on the retrieval performance and the potential effectiveness of our approach, the performance achieved when using the best $\lambda_r$ for each retrieval measure on each test collection (i.e.\ best possible setting) is also reported, denoted \emph{oracle}.


Table~\ref{tab:c7Bayesian} compares the retrieval performances of the Bayesian-based approach with the three baselines on TREC 2011 and 2012 Medical Records track test collection. From Table~\ref{tab:c7Bayesian}, we observe that the Bo1 baseline outperforms the task-specific representation baseline (i.e.\ no QE) for both TREC 2011 and 2012. Specifically, the retrieval performances improve from bpref 0.5243 to 0.5403 and from infNDCG 0.4880 to 0.5129 for TREC 2011 and 2012, respectively. Meanwhile, the Bo1 \& Semantic QE baseline decreases the retrieval performance. 

In contrast, the Bayesian-based approach, using either a 5-fold or an x-collection training regime, outperforms the task-specific representation baseline (both TREC 2011 and 2012). Specifically, the Bayesian-based approach with the 5-fold regime (5-fold) outperforms all of the three used baselines. The Bayesian-based approach (5-fold) achieves a bpref of 0.5474 and an infNDCG of 0.5133. Importantly, our approach (5-fold) significantly (paired t-test, $p<0.05$) outperforms the task-specific representation baseline in terms of infNDCG for TREC 2012 (0.5133 vs 0.4880). 

When comparing the Bayesian-based approach with the Bo1 \& Semantic QE baseline, which also uses the same set of resources as the Bayesian-based approach, we observe that the Bayesian-based approach, either using the 5-fold or x-collection training regime, significantly ($p<0.05$) outperforms this baseline for both the TREC 2011 and 2012 queries. However, the Bayesian-based approach could not significantly outperform the Bo1 baseline.


Additionally, as expected, we find that with a proper setting of the parameter $\lambda_r$ (i.e.\ the $\lambda_r$ that results in the best retrieval performance for each query), the Bayesian-based approach (oracle) can achieve a better retrieval performance. 
In particular, the bpref retrieval performance is increased to 0.5646, while the infNDCG retrieval performance is improved to 0.5393 (+7.69\% and +10.51\% over the task-specific representation baseline, respectively). This shows the potential of the Bayesian-based approach if the $\lambda_r$ value could be effectively set. 


Later, in Section~\ref{c7sAnalysis}, we further analyse the Bayesian-based approach, in comparison with the task-specific representation and the Bo1 baselines.

\subsection{Experiments with a Stochastic Approach for Inferring Conceptual Relationships}\label{c7sExperimentRandomWalk}

Next, we discuss the evaluation of the stochastic approach for inferring the conceptual relationships to improve the representation of a given query. We compare the retrieval effectiveness of our approach with the same baselines as in Section~\ref{c7sExperimentBayesian}.
Table~\ref{tab:c7Salsa} shows the retrieval performances in terms of bpref and infNDCG for TREC 2011 and 2012, respectively.

\begin{table}[tb]
\caption{Retrieval performance of our stochastic approach for inferring relationships between medical concepts in comparison with the three baselines on TREC 2011 and 2012 Medical Records track's queries. Statistical significance (paired t-test) at $p<0.05$ over the task-specific representation and the Bo1 \& Semantic QE baselines are denoted $^t$ and $^{s}$, respectively.}\label{tab:c7Salsa}
  \centering
\resizebox{1.0\textwidth}{!}{
  \begin{tabular}{|l|c|c|}
    \hline
     Approach & bpref (TREC 2011) & infNDCG (TREC 2012) \\ 
    \hline
     Task-specific representation & 0.5243 & 0.4880\\
     + Bo1 & {\bf0.5403} & {\bf0.5129} \\
     + Bo1 \& Semantic QE & 0.3013 & 0.2228\\
     + Our stochastic approach (5-fold) & 0.5355$^{s}$ & 0.4846$^{s}$ \\
     + Our stochastic approach (x-collection) & 0.5376$^{s}$ & 0.4761$^{s}$ \\
\hline
     + Our stochastic approach (oracle) & 0.5586$^{t,s}$ & 0.5311$^{t,s}$ \\
\hline
  \end{tabular}
}
\end{table}

From Table~\ref{tab:c7Salsa}, we observe that the stochastic approach significantly ($p<0.05$) outperforms the Bo1 \& Semantic QE baseline for both TREC 2011 and 2012 queries. In addition, the stochastic approach
outperforms the task-specific representation baseline for the TREC 2011 queries. Specifically, our approach with the 5-fold and the x-collection training regimes achieves bpref 0.5355 and 0.5376, respectively, in comparison with bpref 0.5243 for the task-specific representation baseline. However, for the TREC 2012 queries, our approach (either with the 5-fold or the x-collection training regime) could not improve the retrieval performance over the task-specific representation baseline. Meanwhile, we observe that the Bo1 baseline performs better than the stochastic approach. This is partly due to the limited number of available queries for training the $\lambda_r$ parameter (in Equation~(\ref{eq:combining})). 

When considering the oracle setting with $\lambda_r$ set to the most effective value for each query, we observe that the stochastic approach could further improve the retrieval performance significantly. This shows the potential of the stochastic approach. However, we leave for future work the study of a more effective training of $\lambda_r$.

Next, when comparing the two proposed approaches (see Tables~\ref{tab:c7Bayesian} and \ref{tab:c7Salsa}), we observe that the Bayesian-based approach is more effective than the stochastic approach for both TREC 2011 and TREC 2012. For example, with the 5-fold training regime, the Bayesian-based approach achieves infNDCG 0.5133, while the infNDCG performance of the stochastic approach is 0.4846.



\section{Analysis and Discussion}\label{c7sAnalysis}
This section further discusses the performances of the Bayesian-based and the stochastic approaches for inferring the relationships between medical concepts. 
Specifically, Section~\ref{c7sFailure} discusses which types of queries are likely to benefit from these proposed approaches, while Sections~\ref{c7sAnalysisTask} and~\ref{c7sAnalysisBo1} compare the performances of the two proposed approaches with the task-specific representation baseline and the Bo1 baseline, respectively, on a per-query basis.

\subsection{Failure Analysis}\label{c7sFailure}

In this section, we analyse the performances of both the Bayesian-based and the stochastic approaches in comparison with the task-specific representation baseline to identify which types of queries that are likely to benefit from each of the two approaches. 
Tables~\ref{tab:c7:analysis-bayesian} and~\ref{tab:c7:analysis-salsa} show the numbers of queries benefited or harmed by the Bayesian-based and the stochastic approaches, respectively, which are grouped based on the types of the medical concepts in the queries. 

From Tables~\ref{tab:c7:analysis-bayesian} and~\ref{tab:c7:analysis-salsa}, we observe that both the Bayesian-based approach and the stochastic approach are more likely to be effective when the query contains medical concepts related to diagnostic tests. In particular, the Bayesian-based approach and the stochastic approach improve the retrieval performance for 57.7\% and 55.8\% of the queries containing medical concepts related to diagnostic tests, respectively. 
%
Meanwhile, we observe that the queries that contain medical concepts related to all of the four aspects are less likely to benefit from our two approaches.  

\begin{table}[tb]
\caption{Analysis of our Bayesian-based approach w.r.t.\ the aspects of medical concepts found in the queries. The numbers between the parentheses indicate the number of queries impacted (benefited/harmed) compared to the total number of queries.}\label{tab:c7:analysis-bayesian}
  \centering
\begin{tabular}{|l|c|c|}
    \hline
    Aspects of medical concepts in the query & Benefited & Harmed \\ \hline
    Symptom & 52.9\% (37/70)& 42.9\%(30/70) \\
    Diagnostic test & 57.7\% (30/52) & 38.5\%(20/52)\\
    Diagnosis & 55.9\% (38/68) & 38.2\% (26/68)\\
    Treatment & 55.8\% (29/52) & 40.4\% (21/52)\\
    All 4 & 50\% (16/32) & 46.9\%(15/32) \\
    \hline
  \end{tabular}
\end{table}

\begin{table}[tb]
\caption{Analysis of our stochastic approach w.r.t.\ the aspects of medical concepts found the queries. The numbers between the parentheses indicate the number of queries impacted (benefited/harmed) compared to the total number of queries.}\label{tab:c7:analysis-salsa}
  \centering
  \small
\begin{tabular}{|l|c|c|}
    \hline
    Aspects of medical concepts in the query & Benefited & Harmed \\ \hline
    Symptom & 52.9\% (37/70)& 41.4\%(29/70) \\
    Diagnostic test & 55.8\% (29/52) & 40.4\%(21/52)\\
    Diagnosis & 51.5\% (35/68) & 41.2\% (28/68)\\
    Treatment & 53.8\% (28/52) & 38.5\% (20/52)\\
    All 4 & 50\% (16/32) & 43.8\%(14/32) \\
    \hline
  \end{tabular}
\end{table}

\subsection{Comparison with the Task-Specific Representation Baseline}\label{c7sAnalysisTask}

\begin{figure}[tb]
\centering
        \subfigure[TREC 2011]{
            \label{fig:CR1}
            \includegraphics[angle=270,width=95mm,scale=0.5]{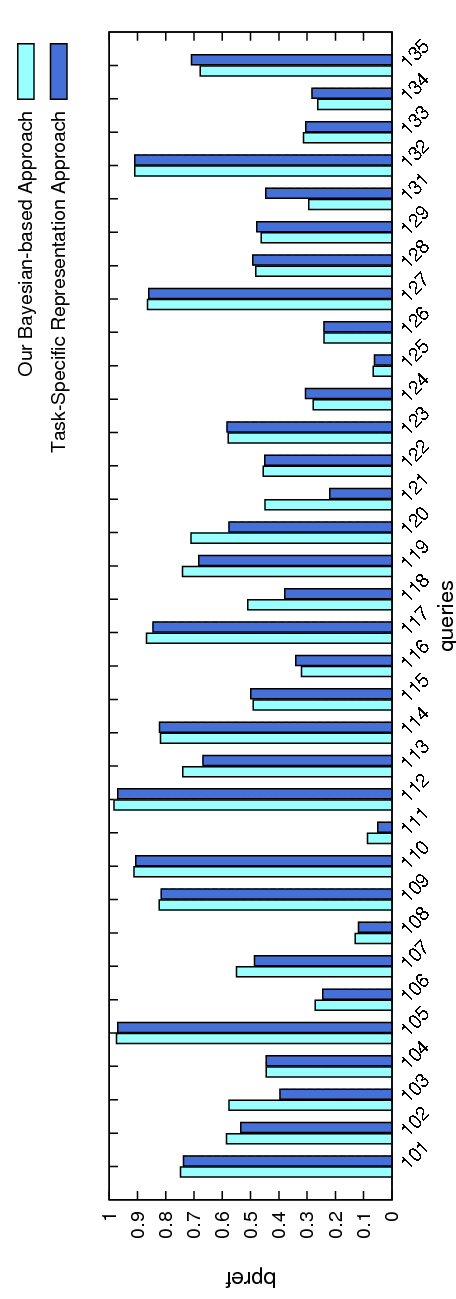}
	}\\%
        \subfigure[TREC 2012]{
           \label{fig:CR2}
           \includegraphics[angle=270,width=95mm,scale=0.5]{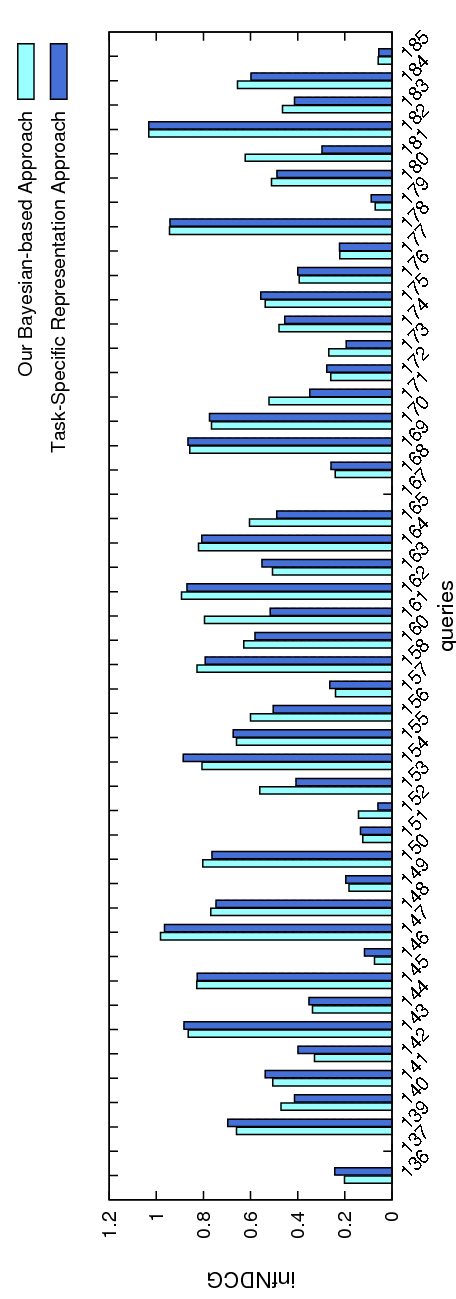}
        } 
    \caption{%
        The retrieval performances of our Bayesian-based approach for inferring the conceptual relationships and the task-specific representation baseline on a per-query basis, evaluated using the TREC 2011 and 2012 Medical Records Track.
     }%
   \label{fig:ch7:bayesian}
\end{figure}

Next, we compare the retrieval performance of the Bayesian-based and the stochastic approaches with the task-specific representation baseline on a per-query basis. We choose to discuss only the 5-fold setting as it is the most effective.

\begin{figure}[tb]
\centering
        \subfigure[TREC 2011]{
            \label{fig:CR1}
            \includegraphics[angle=270,width=95mm,scale=0.5]{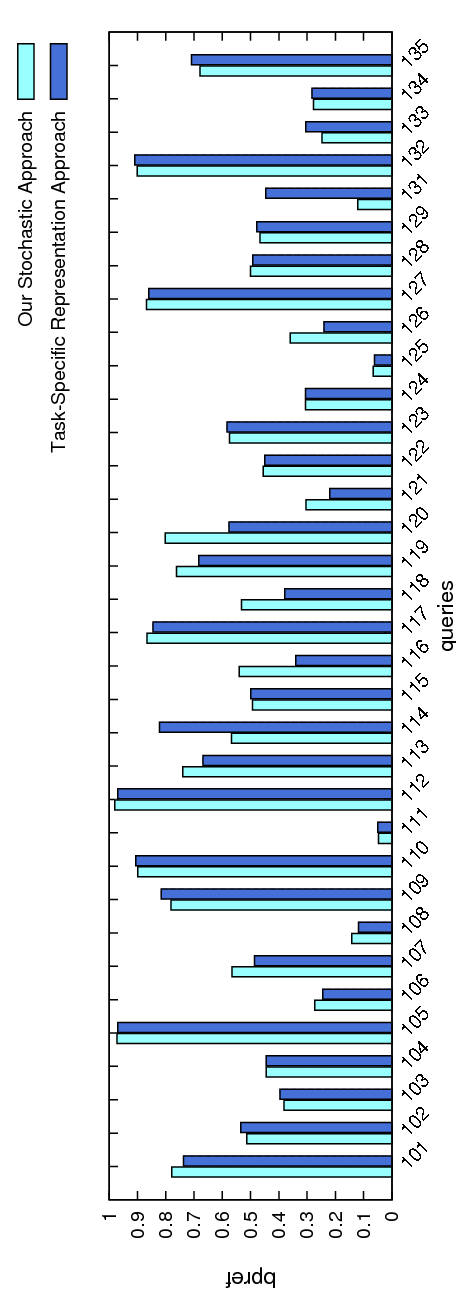}
	}\\%
        \subfigure[TREC 2012]{
           \label{fig:CR2}
           \includegraphics[angle=270,width=95mm,scale=0.5]{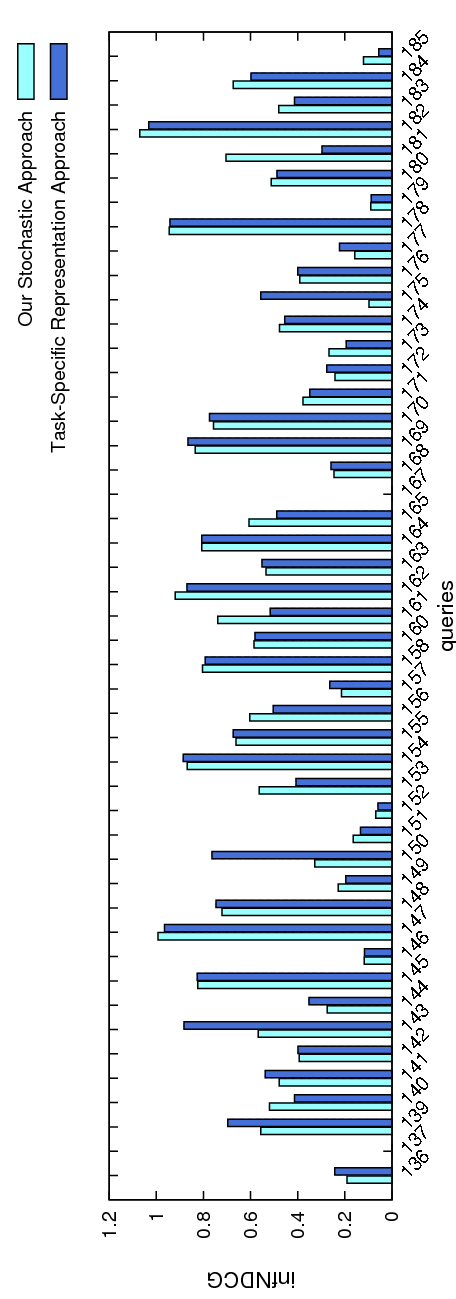}
        } 
    \caption{%
        The retrieval performances of our stochastic approach for inferring the conceptual relationships and the task-specific representation baseline on a per-query basis, evaluated using the TREC 2011 and 2012 Medical Records Track.
     }%
   \label{fig:ch7:salsa}
\end{figure}

We first discuss the performance of the Bayesian-based approach. From Figure~\ref{fig:ch7:bayesian}, we observe that in general the Bayesian-based approach performs better than the task-specific representation baseline. In particular, for the TREC 2011 queries, the Bayesian-based approach outperforms the task-specific representation baseline for 22 out of 34 queries, while it performs worse than the baseline for 10 queries. We observe that the Bayesian-based approach is likely to be more effective for difficult queries. For instance, the Bayesian-based approach could improve the retrieval performance of the queries with a bpref retrieval performance less than 0.25 for 5 out of 6 queries (e.g.\ queries\# 108, 111, 121, and 125). Meanwhile, for the TREC 2012 queries, the Bayesian-based approach outperforms the task-specific representation baseline for 23 out of 47 queries, while performing worse than the baseline for 22 queries.

In Figure~\ref{fig:ch7:salsa}, we show the per-query retrieval effectiveness of the stochastic approach in comparison with the task-specific representation baseline. We observe that the stochastic approach outperforms this baseline for the majority of the queries. Indeed, our approach is more effective for 18 out of 34 queries and for 24 out of 47 queries from TREC 2011 and 2012, respectively. Similar to the Bayesian-based approach, we also observe that the stochastic approach tends to be effective for difficult queries. 
Specifically, for the queries where the task-specific representation baseline obtains a retrieval performance less than 0.25 (bpref for TREC 2011 and infNDCG for TREC 2012), the stochastic approach outperforms the task-specific representation baseline for 7 out of 11 queries for TREC 2011 and for 5 out of 6 for TREC 2012. 

%

\subsection{Comparison with the Bo1 Baseline}\label{c7sAnalysisBo1}


This section discusses the retrieval performances of the Bayesian-based and the stochastic approaches in comparison with the Bo1 baseline, which is a stronger baseline. We compare the retrieval performances on a per-query basis.

\begin{figure}[tb]
\centering
        \subfigure[TREC 2011]{
            \label{fig:CR1}
            \includegraphics[angle=270,width=95mm,scale=0.5]{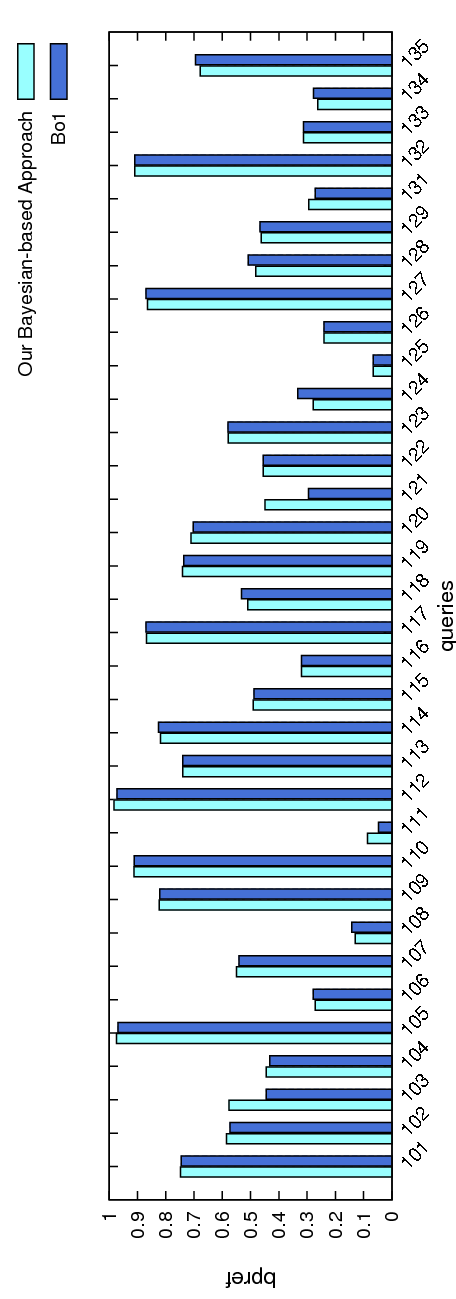}
	}\\%
        \subfigure[TREC 2012]{
           \label{fig:CR2}
           \includegraphics[angle=270,width=95mm,scale=0.5]{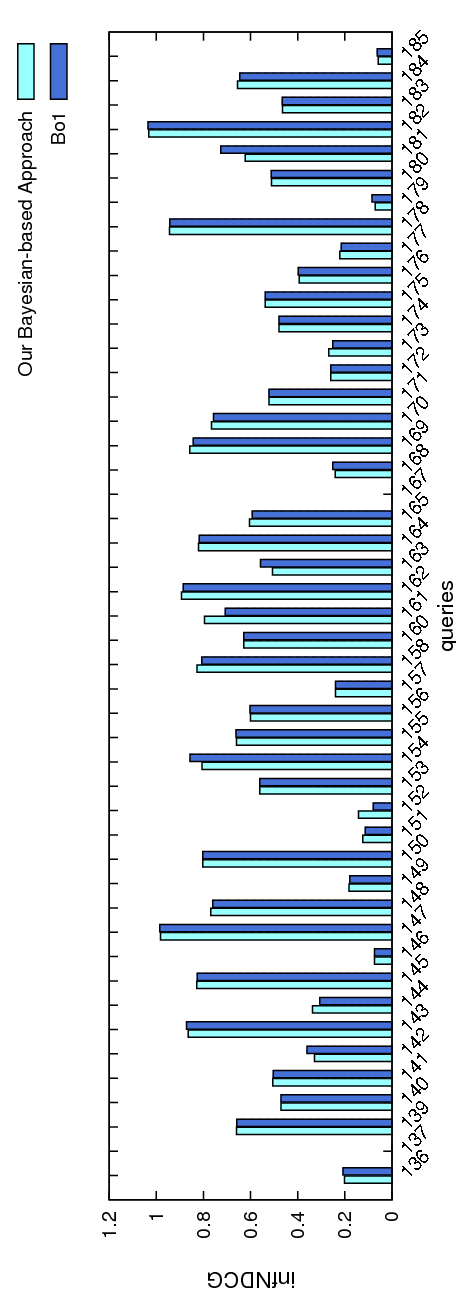}
        } 
    \caption{%
        The retrieval performances of our Bayesian-based approach for inferring the conceptual relationships and the Bo1 query baseline on a per-query basis, evaluated using the TREC 2011 and 2012 Medical Records Track.
     }%
   \label{fig:ch7:bayesian-bo1}
\end{figure}

We first discuss the retrieval effectiveness of the Bayesian-based approach. From Figure~\ref{fig:ch7:bayesian-bo1}, we observe that the Bayesian-based approach performs comparable to the Bo1 baseline. Indeed, our approach outperforms the Bo1 baseline for 16 out of 34 queries and for 21 out of 47 queries for TREC 2011 and TREC 2012, respectively. Meanwhile, the Bo1 baseline performs better than our approach for 12 and 20 queries for TREC 2011 and 2012. 
%

\begin{figure}[tb]
\centering
        \subfigure[TREC 2011]{
            \label{fig:CR1}
            \includegraphics[angle=270,width=95mm,scale=0.5]{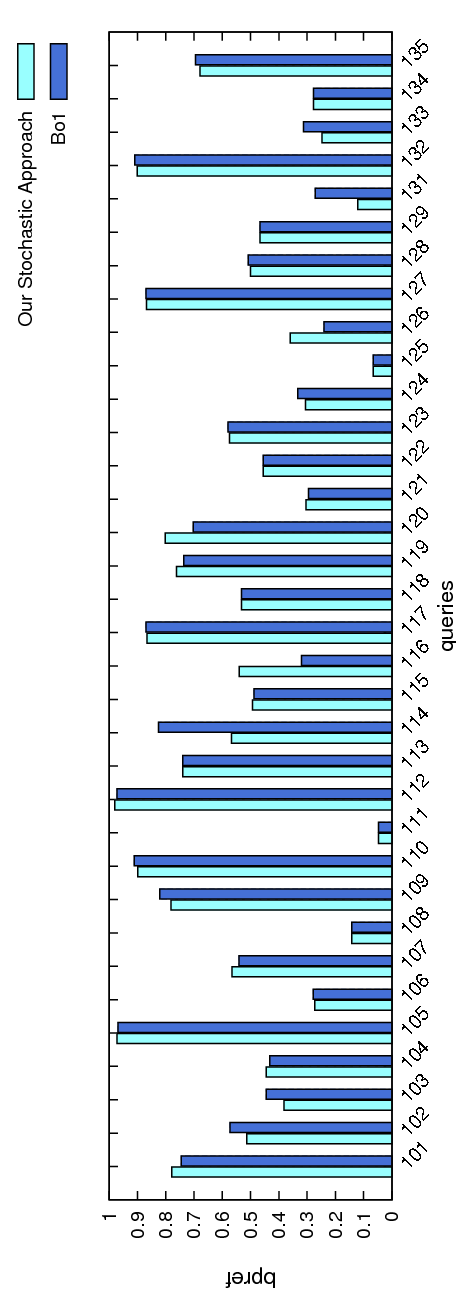}
	}\\%
        \subfigure[TREC 2012]{
           \label{fig:CR2}
           \includegraphics[angle=270,width=95mm,scale=0.5]{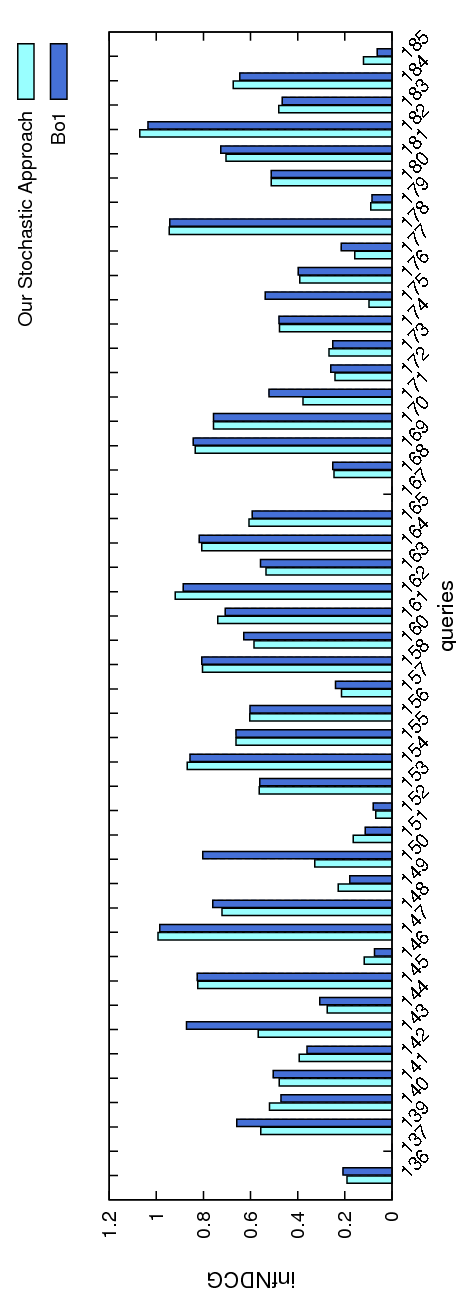}
        } 
    \caption{%
        The retrieval performances of our stochastic approach for inferring the conceptual relationships and the Bo1 baseline on a per-query basis, evaluated using the TREC 2011 and 2012 Medical Records Track.
     }%
   \label{fig:ch7:salsa-bo1}
\end{figure}

Next, we discuss the retrieval performance of the proposed stochastic approach. From Figure~\ref{fig:ch7:salsa-bo1}, we observe that the Bo1 baseline benefits more queries than the stochastic approach. Specifically, the Bo1 baseline outperforms the stochastic approach for 16 (resp.\ 23) out of 34 (resp.\ 47) queries for TREC 2011 (resp.\ TREC 2012), while the stochastic approach performs better for 11 and 20 queries for TREC 2011 and 2012, respectively. However, this is expected, as the Bo1 baseline achieves a better retrieval performance than the stochastic approach, as shown in Table~\ref{tab:c7Salsa}. Nevertheless, when considering the performance on \emph{the difficult queries} (i.e.\ queries that obtain retrieval performance $<$ 0.25 when query expansion is not applied), we observe that the stochastic approach performs better than the Bo1 baseline for 8 out of 16 queries across the TREC 2011 and 2012, while it performs worse than the Bo1 baseline for only 3 queries.


From the per-query performance comparison in Sections~\ref{c7sAnalysisTask} and~\ref{c7sAnalysisBo1}, we observe that both the Bayesian-based and the stochastic approaches are more likely to benefit the difficult queries (i.e.\ the queries that obtain retrieval performance $<$ 0.25 when do not apply query expansion).

\section{Conclusions}\label{Conclusions}

We have proposed a novel approach that uncovers implicit knowledge in medical records and queries. In particular, the proposed approach to infer the relationships between medical conditions by using both external resources (e.g. ontologies and health-related websites) and the local statistics of top-ranked medical records to reformulate the queries. 
In particular, we extracted association rules of relationships between medical concepts from external resources. Then, to infer the relationships of medical concepts from the association rules, we introduced two techniques based on Bayes' theorem and a stochastic analysis, respectively. The medical concepts inferred using the association rules are combined with the medical concepts derived using a pseudo-relevance feedback technique in order to improve the representation of the query.
From the experimental results, we observed that with a fair setting the Bayesian-based approach improved the retrieval performance by up to 10.51\% over the baseline that does not apply a QE technique. Meanwhile, the stochastic approach was likely to perform effectively for difficult queries (i.e.\ the queries where the retrieval performance of the original queries are less than 0.25 in term of bpref and infNDCG for TREC 2011 and 2012, respectively). In addition, we observed that both the Bayesian-based and the stochastic approaches were less likely to be effective for the queries that contain medical concepts related to all of the four aspects of the medical decision process. For future work, we aim to investigate a more effective technique for extracting association rules from different external resources and compare which resources are more useful for the patient search task.

\bibliographystyle{unsrt}
\bibliography{limsopatham2014medicalir}

\begin{thebibliography}{10}

\bibitem{kotsipoulos}
I.~A. Kotsiopoulos, J.~A. Keane, M.~Turner, P.~J. Layzell, and F.~Zhu.
\newblock Ibhis: Integration broker for heterogeneous information sources.
\newblock In {\em Proceedings of the 27th Annual International Computer
  Software and Applications Conference}, 2003.

\bibitem{tambouris}
E.~Tambouris and C.~Makropoulos.
\newblock Hin6/427: Co-operative health information network in europe: The
  greek experience.
\newblock {\em Journal of Medical Internet Research}, 2(2), Sep 1999.

\bibitem{limsopatham2013oair}
N.~Limsopatham, C.~Macdonald, and I.~Ounis.
\newblock Inferring conceptual relationships to improve medical records search.
\newblock In {\em Proceedings of the 10th Open research Areas in Information
  Retrieval}, Lisbon, Portugal, 2013.

\bibitem{voorhees2011trec}
E.~Voorhees and R~Tong.
\newblock Overview of the {TREC} 2011 medical records track.
\newblock In {\em Proceedings of the 20th Text REtrieval Conference},
  Gaithersburg, MD, USA, 2011.

\bibitem{voorhees2012trec}
E.~Voorhees and W~Hersh.
\newblock Overview of the {TREC} 2012 medical records track.
\newblock In {\em Proceedings of the 21st Text REtrieval Conference},
  Gaithersburg, MD, USA, 2012.

\bibitem{krauthammer2004jbi}
Michael Krauthammer and Goran Nenadic.
\newblock Term identification in the biomedical literature.
\newblock {\em J. of Biomedical Informatics}, 37(6):512--526, December 2004.

\bibitem{limsopatham2011sigir}
N.~Limsopatham, R.~L.~T. Santos, C.~Macdonald, and I.~Ounis.
\newblock Disambiguating biomedical acronyms using emim.
\newblock In {\em Proceedings of the 34th Annual International ACM SIGIR
  Conference on Research and Development in Information Retrieval}, Beijing,
  China, 2011.

\bibitem{trieschnigg2010cikm}
D.~Trieschnigg, D.~Hiemstra, F.~de~Jong, and W.~Kraaij.
\newblock A cross-lingual framework for monolingual biomedical information
  retrieval.
\newblock In {\em Proceedings of the 19th ACM International Conference on
  Information and Knowledge Management}, New York, NY, USA, 2010.

\bibitem{hersh1994jamia}
W.~Hersh, D.~Hickam, R.~Haynes, and K.~McKibbon.
\newblock A performance and failure analysis of saphire with a medline test
  collection.
\newblock {\em Journal of the American Medical Informatics Association}, 1(1),
  1994.

\bibitem{srinivasan1996ipma}
P.~Srinivasan.
\newblock Optimal document-indexing vocabulary for medline.
\newblock {\em Information Processing and Management}, 32(5), September 1996.

\bibitem{aronson1997amia}
A.~R. Aronson and T.~C. Rindflesch.
\newblock Query expansion using the {UMLS} {M}etathesaurus.
\newblock {\em Proceedings of the American Medical Informatics Association
  (AMIA) Symposium}, 1997.

\bibitem{srinivasan1996ipmb}
Padmini Srinivasan.
\newblock Query expansion and medline.
\newblock {\em Inf. Process. Manage.}, 32(4):431--443, July 1996.

\bibitem{amati2003thesis}
G.~Amati.
\newblock {\em Probabilistic Models for Information Retrieval based on
  Divergence from Randomness}.
\newblock PhD thesis, University of Glasgow, 2003.

\bibitem{limsopatham2012sigir}
N.~Limsopatham, C.~Macdonald, R.~McCreadie, and I.~Ounis.
\newblock Exploiting term dependence while handling negation in medical search.
\newblock In {\em Proceedings of the 35th International ACM SIGIR Conference on
  Research and Development in Information Retrieval}, New York, NY, USA, 2012.

\bibitem{limsopatham2015cikm}
Nut Limsopatham, Craig Macdonald, and Iadh Ounis.
\newblock Modelling the usefulness of document collections for query expansion
  in patient search.
\newblock In {\em Proceedings of the 24th ACM International on Conference on
  Information and Knowledge Management}, CIKM '15, pages 1739--1742, New York,
  NY, USA, 2015. ACM.

\bibitem{Silfen2006664}
E.~Silfen.
\newblock Documentation and coding of ed patient encounters: an evaluation of
  the accuracy of an electronic medical record.
\newblock {\em The American Journal of Emergency Medicine}, 24(6), 2006.

\bibitem{limsopatham2013ecir-b}
N.~Limsopatham, C.~Macdonald, and I.~Ounis.
\newblock A task-specific query and document representation for medical records
  search.
\newblock In {\em Proceedings of the 35th European Conference on Information
  Retrieval}, Moscow, Russia, 2013.

\bibitem{hersh2009bio}
W.~Hersh.
\newblock {\em {Information Retrieval: A Health and Biomedical Perspective
  (Health Informatics)}}.
\newblock Springer, 3rd edition, November 2008.

\bibitem{hersh2004jama}
W.~Hersh.
\newblock Health care information technology: Progress and barriers.
\newblock {\em JAMA: The Journal of the American Medical Association}, 292(18),
  2004.

\bibitem{gurulingappa2011trec}
H.~Gurulingappa, B.~Uller, M.~Hofmann-Apitius, and J.~Fluck.
\newblock A semantic platform for information retrieval from e-health records.
\newblock In {\em Proceedings of the 20th Text REtrieval Conference},
  Gaithersburg, MD, USA, 2011.

\bibitem{limsopatham2011trec}
N.~Limsopatham, C.~Macdonald, I.~Ounis, G.~McDonald, and M.~Bouamrane.
\newblock University of glasgow at medical records track 2011: Experiments with
  terrier.
\newblock In {\em Proceedings of the 20th Text REtrieval Conference},
  Gaithersburg, MD, USA, 2011.

\bibitem{zhu2012shb}
D.~Zhu and B.~Carterette.
\newblock Combining multi-level evidence for medical record retrieval.
\newblock In {\em Proceedings of the 2012 International Workshop on Smart
  Health and Wellbeing}, New York, NY, USA, 2012.

\bibitem{balog2008trec}
K.~Balog, P.~Thomas, N.~Craswell, I.~Soboroff, , and P.~Bailey.
\newblock Overview of the trec 2008 enterprise track.
\newblock In {\em Proceedings of the 17th Text REtrieval Conference},
  Gaithersburg, MD, USA, 2008.

\bibitem{macdonald2006cikm}
C.~Macdonald and I.~Ounis.
\newblock Voting for candidates: adapting data fusion techniques for an expert
  search task.
\newblock In {\em Proceedings of the 15th ACM international conference on
  Information and knowledge management}, New York, NY, USA, 2006.

\bibitem{balog2006sigir}
K.~Balog, L.~Azzopardi, and M.~de~Rijke.
\newblock Formal models for expert finding in enterprise corpora.
\newblock In {\em Proceedings of the 29th Annual International ACM SIGIR
  Conference on Research and Development in Information Retrieval}, Seattle,
  Washington, USA, 2006.

\bibitem{robertson1994sigir}
S.~E. Robertson and S.~Walker.
\newblock Some simple effective approximations to the 2-poisson model for
  probabilistic weighted retrieval.
\newblock In {\em Proceedings of the 17th Annual International ACM SIGIR
  Conference on Research and Development in Information Retrieval}, New York,
  NY, USA, 1994.

\bibitem{amati2007trec}
G.~Amati, E.~Ambrosi, M.~Bianchi, C.~Gaibisso, and G.~Gambosi.
\newblock Fub, iasi-cnr and university of tor vergata at trec 2007 blog track.
\newblock In {\em Proceedings of the 16th Text REtrieval Conference},
  Gaithersburg, MD, USA, 2007.

\bibitem{aronson1994riao}
A.~R. Aronson.
\newblock Exploiting a large thesaurus for information retrieval.
\newblock In {\em Proceedings of RIAO 1994}, 1994.

\bibitem{aronson2010jamia}
A.~R. Aronson and F.~M. Lang.
\newblock An overview of metamap: historical perspective and recent advances.
\newblock {\em Journal of the American Medical Informatics Association}, 17(3),
  2010.

\bibitem{koopman2012adcs}
B.~Koopman, G.~Zuccon, P.~Bruza, L.~Sitbon, and M.~Lawley.
\newblock Graph-based concept weighting for medical information retrieval.
\newblock In {\em Proceedings of the Seventeenth Australasian Document
  Computing Symposium}, New York, NY, USA, 2012.

\bibitem{brin1998}
S.~Brin and L.~Page.
\newblock The anatomy of a large-scale hypertextual web search engine.
\newblock In {\em Proceedings of the Seventh International Conference on World
  Wide Web 7}, Amsterdam, The Netherlands, The Netherlands, 1998.

\bibitem{koopman2014sigir}
B.~Koopman and G.~Zuccon.
\newblock Understanding negation and family history to improve clinical
  information retrieval.
\newblock In {\em Proceedings of the 37th International ACM SIGIR Conference on
  Research and Development in Information Retrieval}, New York, NY, USA, 2014.

\bibitem{limsopatham2014cikm}
N.~Limsopatham, C.~Macdonald, and I.~Ounis.
\newblock Modelling relevance towards multiple inclusion criteria when ranking
  patients.
\newblock In {\em Proceedings of the 23rd ACM International Conference on
  Information and Knowledge Management}, Shanghai, China, 2014.

\bibitem{martinez2014}
David Martinez, Arantxa Otegi, Aitor Soroa, and Eneko Agirre.
\newblock Improving search over electronic health records using umls-based
  query expansion through random walks.
\newblock {\em Journal of Biomedical Informatics}, 51:100 -- 106, 2014.

\bibitem{zuccon2012}
Guido Zuccon, Bevan Koopman, Anthony Nguyen, Deanne Vickers, and Luke Butt.
\newblock Exploiting medical hierarchies for concept-based information
  retrieval.
\newblock In {\em Proceedings of the Seventeenth Australasian Document
  Computing Symposium}, pages 111--114. ACM, 2012.

\bibitem{king2011trec}
B.~King, L.~Wang, I.~Provalov, and J.~Zhou.
\newblock Cengage learning at trec 2011 medical track.
\newblock In {\em Proceedings of the 20th Text REtrieval Conference},
  Gaithersburg, MD, USA, 2011.

\bibitem{lempel2001tois}
R.~Lempel and S.~Moran.
\newblock Salsa: The stochastic approach for link-structure analysis.
\newblock {\em ACM Transactions on Information Systems}, 19(2), April 2001.

\bibitem{blanco2012ir}
R.~Blanco and C.~Lioma.
\newblock Graph-based term weighting for information retrieval.
\newblock {\em Information Retrieval}, 15(1), February 2012.

\bibitem{tyree2011www}
S.~Tyree, K.~Q. Weinberger, K.~Agrawal, and J.~Paykin.
\newblock Parallel boosted regression trees for web search ranking.
\newblock In {\em Proceedings of the 20th International Conference on World
  Wide Web}, New York, NY, USA, 2011.

\bibitem{ganjisaffar2011sigir}
Y.~Ganjisaffar, R.~Caruana, and C.~V. Lopes.
\newblock Bagging gradient-boosted trees for high precision, low variance
  ranking models.
\newblock In {\em Proceedings of the 34th International ACM SIGIR Conference on
  Research and Development in Information Retrieval}, New York, NY, USA, 2011.

\bibitem{limsopatham2013sigir}
N.~Limsopatham, C.~Macdonald, R.~McCreadie, and I.~Ounis.
\newblock Learning to combine representations for medical records search.
\newblock In {\em Proceedings of the 36th Annual International ACM SIGIR
  Conference on Research and Development in Information Retrieval}, Dublin,
  Ireland, 2013.

\bibitem{cronen-townsend2002sigir}
S.~Cronen-Townsend, Y.~Zhou, and W.~B. Croft.
\newblock Predicting query performance.
\newblock In {\em Proceedings of the 25th Annual International ACM SIGIR
  Conference on Research and Development in Information Retrieval}, New York,
  NY, USA, 2002.

\bibitem{zhao2008ecir}
Y.~Zhao, F.~Scholer, and Y.~Tsegay.
\newblock Effective pre-retrieval query performance prediction using similarity
  and variability evidence.
\newblock In {\em Proceedings of the IR Research, 30th European Conference on
  Advances in Information Retrieval}, Berlin, Heidelberg, 2008.

\bibitem{carmel2010sl}
D.~Carmel and E.~Yom-Tov.
\newblock {\em Estimating the Query Difficulty for Information Retrieval}.
\newblock Synthesis Lectures on Information Concepts, Retrieval, and Services.
  Morgan {\&} Claypool Publishers, 2010.

\bibitem{he2006is}
B.~He and I.~Ounis.
\newblock Query performance prediction.
\newblock {\em Information Systems}, 31(7), November 2006.

\bibitem{buckley2004sigir}
C.~Buckley and E.~M. Voorhees.
\newblock Retrieval evaluation with incomplete information.
\newblock In {\em Proceedings of the 27th Annual International ACM SIGIR
  Conference on Research and Development in Information Retrieval}, New York,
  NY, USA, 2004.

\bibitem{yilmaz2008sigir}
E.~Yilmaz, E.~Kanoulas, and J.~A. Aslam.
\newblock A simple and efficient sampling method for estimating ap and ndcg.
\newblock In {\em Proceedings of the 31st Annual International ACM SIGIR
  Conference on Research and Development in Information Retrieval}, New York,
  NY, USA, 2008.

\bibitem{ounis06terrier-osir}
I.~Ounis, G.~Amati, V.~Plachouras, B.~He, C.~Macdonald, and C.~Lioma.
\newblock {Terrier: A High Performance and Scalable Information Retrieval
  Platform}.
\newblock In {\em Proceedings of ACM SIGIR'06 Workshop on Open Source
  Information Retrieval (OSIR 2006)}, 2006.

\end{thebibliography}


\end{document}